# The Dynamics of the intermittency maps reveal the existence of resonances phenomena, interesting hybrid states and the orders of the phase transitions in a finite Z(3) spin model in 3D Lattice.

Yiannis F. Contoyiannis

Department of Electrical and Electronics Engineering, University of West Attica, Ancient Olive Grove Campus, 250 Thivon and P. Ralli, Athens GR12244, Greece.

*Correspondence: yiaconto@uniwa.gr

Abstract : A numerical simulation using the chaotic Dynamics of intermittency at a finite size Z(3) spin system in a 3D lattice reveals: (a) the existence of a second order phase transition with a zone hysteresis characterized from resonances phenomena (b) An hybrid appearance of mean-field universality class and 3D Ising model universality class , all these inside the zone hysteresis (c) a weak first order phase transition through a tricritical crossover. So, a complicated behavior in Z(3) symmetry exists.

## 1. Introduction

The Z(3) spin model belongs to the Z(N) models and has been studied extensively in the past. For a Z(N) spin system spin variables are defined by $s(a_i) = e^{j2\pi a_i/N}$ with $a_i = 0,1,2,3 \ldots N-1$. The action for the spin system possessing a global Z(N) symmetry can be written as [1,2]:

$$S_E = -\beta_B J \sum_{\{1,2\}} Re\{s(-a_1)s(a_2)\} = -\beta_B J \sum_{\{1,2\}} Re\{s^*(a_1)s(a_2)\} =$$

$$-\beta_B J \sum_{\{1,2\}} \cos\{\frac{2\pi}{N}(a_2 - a_1)\} \quad (1)$$

where {1,2} labels a nearest-neighboring pair and the summation is performed over the whole set of nearest-neighboring pair in the lattice. The coefficient $\beta_B$ is the inverse temperature $\beta_B = \frac{1}{T}$ and J>0 is the strength of the interactions between the spins in Lattice sites . Therefore, for N=3 the values of $a_1, a_2$ are 0,1,2 and the sum terms in equation (1) can take two possible values that equal to 1 when $a_1 = a_2$ i.e when the spins of the nearest-neighbor pair aligned in the same direction , and $-\frac{1}{2}$ when $\alpha_1 \neq \alpha_2$.

The most important question about the Z(N) models is the problem of the phase transition between an absolute symmetrical phase and a phase of symmetry breaking. In the simplest Z(2) model where two orientation for spin vector exist (up and down) which is known as Ising models( 3D or 2D) the phase transition is a continuous transition i. e a second order phase transition. When the orientations of spin vector increases is expected that complexity about the phase transition will also increase. Thus a great number of models Z(3) has been development . The most works have results that Z(3) spin models in the 3D space in the framework of Mean field approximation, show a first order phase transition [3,4]. In reference [5] the authors use a method to study the phase transition of the three-dimensional Z ( 3 ) spin model with nearest neighbors and next nearest couplings . The system shows a rich phase diagram [22] when considered the existence of two coupling each for every phase. The results are that in this phase diagram a first-order phase transition is appears but in a region of diagram a second order phase transition exist too. The approximation about the nearest neighbors and next nearest couplings used in work [6] for the triangular Ising model of finite size show that a weak first order phase transition appears too. In general, we could say that the type of phase transition in Z(N) spin models is significantly influenced by the coupling strengths between lattice sites, the lattice geometry, the addition of next nearest couplings, the choice of order parameters [7] and other factors. Z(3) models (or more generally Potts models) are also useful for studying matter at enormous energies through the SU(3) symmetry of QCD, which belongs to the same universality class as the Z(3) model [1] . It is important to know under what conditions one could "encounter", within the matter of the plasma (QGP), continuous transitions (second order phase transitions),or abrupt transitions (first order phase transitions), or resonance phenomena [23] , or a crossover around the tricritical point in the parametric space of the first order phase transition [24].

Our approximation about the issue of the phase transition in a Z(3) spin system , which is the object of the present work, is very different from the up to now works. In an old work of ours [8] we had transfer the notion of the critical point of Statistical

Physics into the notion of critical state which is a Dynamical temporal procedure. This is accomplished through the renormalization group of Hu and Rudnick [9] which is referred at the intermittency maps with a weak chaos [8]. In this way we introduce the temporal fluctuations of Statistical quantities, like order parameter, in critical phenomena. Using the Method of Critical Fluctuations (MCF) [10] which reveals quantitatively the critical state of a second order phase transition, we find that our Z(3) spin model demonstrates a degeneration of its critical point. On the other hand, this degeneration cause an hysteresis zone. Similar hysteresis zone we have found recently in the Spontaneous Symmetry Breaking of Z(2) symmetry which describe the second order phase transition [10,11,12]. Inside this hysteresis zone we find signature from two different Universality class which are the mean-field theory and the 3D Ising model. Moreover we reveal, inside the zone, a resonance phenomenon.

Introducing a different order parameter the signature from a weak first order phase transition appears too. In this way, we confirm the complex behavior and we add new properties of Z(3) symmetry which are unknown up today.

## 2. The phase diagram of the our model

A numerical simulation in a 3D lattice based on Metropolis algorithm and the above action (1) could be accomplished. The structure of our model is as following: Based on the previous we examine a Z(3) model in a 3D cubic lattice with length L. In every vertex (site) of the lattice, spin could have one orientation between three orientations forming an angle of 2π/3 between them and these orientations lies in parallel planes vertically to axis z of cubic lattice. If we name the three orientations as 1, 2,3 then in symmetric phase i.e in high temperatures above the pseudocriticall temperature in the whole cubic lattice the sites have measures of vectors as : 33,33% in 1, 33.33 % in 2 and 33,33 % in 3 ( see table 1). Thus the vectors sum of the spins over the whole lattice is zero. In the phase of broken symmetry below the pseudocritical temperature the measures of spin directions is not the same. So the vector sum of spins, in whole lattice, now it is not zero.

The Monte Carlo simulation is carried out for k iterations for each temperature, producing in this way configurations inside lattice. The input data are : A 3D lattice with L=10, J=1, k=100000 iterations where they have gone before 20000 iterations as thermalization. The procedure has as following :

For each configuration we calculate the magnitude of the sum for vectors $M_{conf} = |f_1 + f_2 + f_3|$, where the vectors $f_1, f_2, f_3$

(on the three orientations) have magnitudes $f_1$, $f_2$, $f_3$ corresponding.

We calculate the statistical weights $w_1$, $w_2$, $w_3$, from $f_1$, $f_2$, $f_3$ values over all k configurations. For each temperature we write the corresponding $w_1$, $w_2$, $w_3$ values on the table I.

Finally, we calculate for each temperature as scalar quantity the order parameter M as the mean value of the fluctuations $M_{conf}$ over all k configurations (iterations), $M = \frac{\sum_{i=1}^{k} M_{conf,i}}{k}$.

Indeed, the quantity M, by its definition, has the characteristics of an order parameter, namely zero in symmetric phase where the statistical weights $w_1$, $w_2$, $w_3$ have the same value and non zero in the phase of symmetry breaking. We repeat the above procedure as the temperature drops and we record the mean values over all iterations of magnitude $w_1$, $w_2$, $w_3$ as well as the order parameter M until all the spin vectors finally will have only a common direction i.e the symmetry breaking has been finished. Thus results the following table I.

## TABLE 1

| T | $w_1$ | $w_2$ | $w_3$ | M | DW |
|---|---|---|---|---|---|
| 6 | 0.333 | 0.333 | 0.333 | 0.038 | 0 |
| 5 | 0.333 | 0.333 | 0.333 | 0.042 | 0 |
| 4 | 0.333 | 0.333 | 0.333 | 0.05 | 0 |
| 3 | 0.333 | 0.333 | 0.333 | 0.09 | 0 |
| 2.9 | 0.333 | 0.333 | 0.333 | 0.11 | 0 |
| 2.8 | 0.333 | 0.334 | 0.332 | 0.16 | 0.001 |
| 2.79 | 0.332 | 0.334 | 0.333 | 0.183 | 0.001 |
| 2.78 | 0.333 | 0.343 | 0.338 | 0.20 | 0.005 |
| 2.77 | 0.329 | 0.319 | 0.34 | 0.22 | 0.01 |
| 2.76 | 0.337 | 0.338 | 0.324 | 0.248 | 0.013 |
| 2.75 | 0.328 | 0.334 | 0.337 | 0.272 | 0.006 |
| 2.74 | 0.331 | 0.335 | 0.334 | 0.325 | 0.003 |
| 2.73 | 0.315 | 0.338 | 0.346 | 0.373 | 0.023 |
| 2.72 | 0.323 | 0.328 | 0.35 | 0.438 | 0.005 |
| 2.71 | 0.321 | 0.297 | 0.382 | 0.494 | 0.024 |
| 2.70 | 0.228 | 0.412 | 0.359 | 0.533 | 0.131 |
| 2.69 | 0.448 | 0.292 | 0.259 | 0.579 | 0.033 |
| 2.68 | 0.626 | 0.204 | 0.170 | 0.607 | 0.034 |
| 2.67 | 0.405 | 0.357 | 0.238 | 0.631 | 0.119 |
| 2.66 | 0.766 | 0.116 | 0.116 | 0.65 | 0 |
| 2.65 | 0.778 | 0.111 | 0.111 | 0.668 | 0 |
| 2.6 | 0.819 | 0.091 | 0.091 | 0.728 | 0 |
| 2.2 | 0.028 | 0.943 | 0.028 | 0.915 | 0 |
| 2 | 0.016 | 0.967 | 0.016 | 0.9513 | 0 |
| 1.5 | 0.003 | 0.994 | 0.003 | 0.991 | 0 |
| 1.35 | 0.002 | 0.996 | 0.002 | 0.996 | 0 |

In this table the values of statistical weights have been rounded to the third decimal place. The last column refers to quantity DW which is the absolute difference between the two smaller statistical weights.

From the above table we conclude the following :

- There is the phase of global symmetry ( or the high temperature phase) for $T > 2.79$ where statistical weights have magnitude equal between them and so the order parameter M is zero . The quantity DW has the value zero too.
- There is the broken symmetry phase for T<2.67 where the global symmetry has been broken but a partial symmetry between the two smaller statistical weights survives until for very small temperatures a only statistical weight dominates . The quantity DW has the value zero.

- Between the two phase an intermediate temperatures zone where $DW \neq 0$ appears (yellow –red zone). In this table we have notice the temperature T=2.79 as a separation line between the high symmetry phase and the DW fluctuations zone. If this zone of fluctuations does not existed this temperature would be the critical point between the global symmetry phase and the broken symmetry phase. Therefore in finite systems a zone hysteresis appears between the symmetric phase and the phase of broken symmetry. As we has wrote in Introduction this phenomenon has been appears in the Z(2) ( Ising models ) too for finite systems . Thus in a finite Z(2) model a result of the hysteresis zone is the displacement of spontaneous symmetry breaking ( SSB) at lower temperatures from the theoretical its values [13].

Based on table 1 we show in figure 1 the phase diagram M vs T.

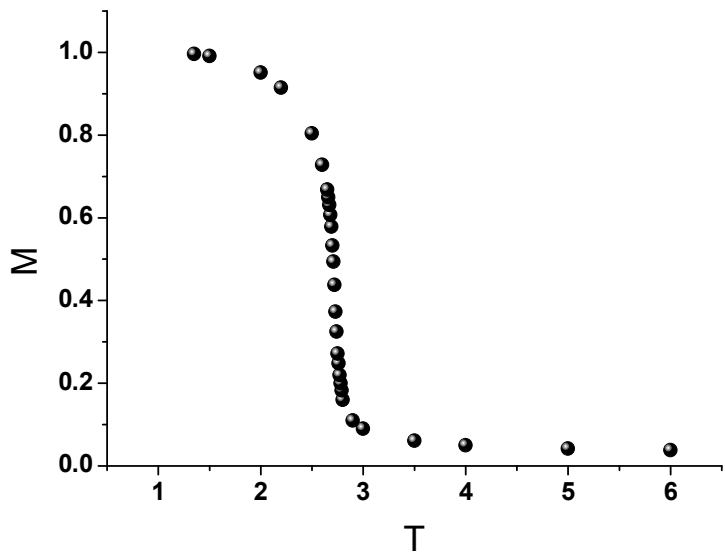

*Fig 1. The phase diagram of the order parameter M vs T .*

From fig.1 results that the evolution of the order parameter M vs T present the quality characteristics of a phase diagram of second order transition [14]. The quantitative characteristics of the second order phase transition will be revealed in following through the application of Method of Critical Fluctuations (MCF).

## 3. The analysis by the Method of Critical fluctuations ( MCF).

As we refer in Introduction in order to estimate the characteristic of critical state , like the critical exponent , the scale-invariance behavior, how close or far a system is to its criticality,.. we use a dynamical method i.e the MCF based on the chaotic Dynamics of intermittency timeseries.

Importantly, the exact dynamics at the critical state can be determined analytically for a large class of critical systems ( artificial or Natural) introducing the so-called critical map. This map can be approximated as an intermittent map type I [**16**]:

$$\phi_{n+1} = \phi_n + u\phi_n^z + \varepsilon_n \qquad (2)$$

The shift parameter $\varepsilon_n$ introduces a non-universal stochastic noise which is necessary for the establishment of ergodicity [**17**]. Each physical system has its characteristic "noise", which is expressed through the shift parameter $\varepsilon_n$. Notice, for thermal systems the exponent z is connected with the isothermal critical exponent $\delta$ as $z = \delta + 1$ [16]. The crucial observation in this approach is the fact that the distribution $P(l)$ of the suitable defined laminar lengths $l$ (waiting times in laminar region) of the above mentioned intermittent map of Eq. (2) in the limit $\varepsilon_n \to 0$ is given by the power law [**17**]

$$P(l) \sim l^{-p_l} \qquad (3)$$

where the exponent $p_l$ is connected with the exponent $z$ by $p_l = \frac{z}{z-1}$. Therefore the exponent $p_l$ is connected with the isothermal exponent $\delta$ by: $p_l = 1 + \frac{1}{\delta}$. The main aim of the MCF is to calculate the exponent $p_l$. We produce the laminar lengths $l$ in a time-series as the lengths that are resulted from successive $\phi$-values obeying the condition $\phi_o \leq \phi \leq \phi_l$, where $\phi_o$ is a marginally stable fixed point and $\phi_l$ is the end of laminar region. In others words the laminar lengths are the waiting times inside the laminar region. The distribution of the laminar lengths (as defined above) is fitted by the function:

$$P(l) \sim l^{-p_2} e^{-p_3 l} \qquad (3)$$

where the exponents $p_2$ and $p_3$ attain a direct physical interpretation as we will discuss in the following. If the exponent $p_3$ is zero, then, the exponent $p_2$ is equal to the exponent $p_l$. The relation $p_l = \frac{z}{z-1}$ suggests that the exponent $p_l$ (or $p_2$) should be greater than 1. On the other hand, from the theory of critical phenomena [**14**] it results that the isothermal exponent $\delta$ is greater than 1. So, from the relation $p_l = 1 + \frac{1}{\delta}$ we obtain the condition $1 < p_l (= p_2) < 2$. In conclusion, the critical profile of the temporal fluctuations is restored for the conditions: $1 < p_l (= p_2) < 2$ and $p_3 \approx 0$. As the system departs from the critical state, then, the exponent $p_2$ decreases while simultaneously the exponent $p_3$ increases reinforcing, in this way, the great exponent cut the long laminar lengths and destroy the scaling laws.

In the table I we had noticed that the temperature T=2.79 is the pseudocritical point of a second order phase transition. Then the fluctuations time-series of the order parameter M must demonstrate critical behavior according to MCF analysis. In the mean field approximation the order of the phase transition depends on the form of action [1]. Due to the fact that a Z(3) spin system has the action of eq.(1) the model present a second order phase transition [1] and belongs to mean field universality class where δ=3 [14]. Therefore from the relation $p_l = 1 + \frac{1}{\delta}$ we must obtain that the expected value of the exponent $p_2$ would be $p_2$ =1+1/3= 1.33.

In figure 2 the application of MCF analysis in the order parameter M at T=2.79 is shown. In figure caption we describe the application of MCF.

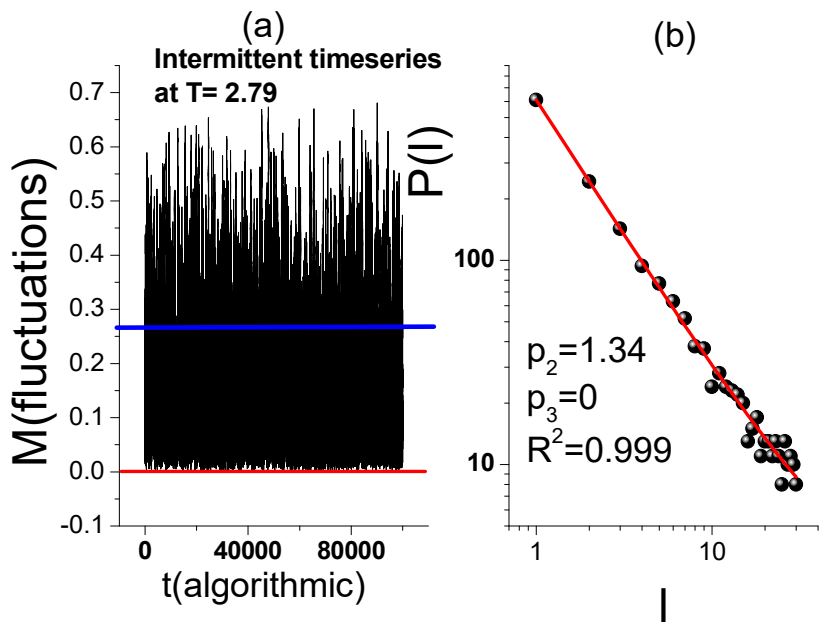

*Figure.2 (a) The timeseries of the order parameter produced by the Monte Carlo algorithm consisted from 100000 points in temperature T=2.79 is shown. The two line (red and blue) determined the boundaries of the laminar region. The red line correspond in fixed point zero and the blue line is a free parameter correspond in the end of laminar region. Based on the condition $M_o \leq M \leq M_l$, where $M_o$ is a marginally stable fixed*

point and $M_l$ is the end of laminar region we record the laminar lengths l. *We had defined as $M_l$ the value of laminar end, where the distribution of laminar lengths is closer to the power-law i.e the parameter is more close to zero i.e $p_3 \approx 0$. In T=2.79 we found that $M_l = 0.27$. Thus the laminar region is [0,0.27]. (b) The laminar lengths distribution produced in laminar region [0,0.27] is shown. In the same diagram the fitting function eq.3 is present too. The quality of power-law is excellent ( $R^2 = 0.999$) and the estimated value are $p_2 = 1.34$ ( very close to the 1.33 of mean field ) and $p_3 \approx 0$.*

The calculation of critical exponent $p_2$ shows, that the exponent has value very close to 1.33. This is the signature about the second order phase transition for a system belonging to the mean field universality class.

### 4. The Degeneration of critical point

In this section we extend the MCF analysis in the all temperatures which are inside hysteresis zone ( table 1). The reason for something like this is what is valid about the criticality inside the whole zone ? Thus, we repeat the procedure of MCF as presented in figure capture in fig.2. Now, in fig3 we present the results about the $p_2$ values for the distributions of laminar region in each temperatures. The $p_3$-values in all cases are found very close to zero. So, all the distributions are power-law.

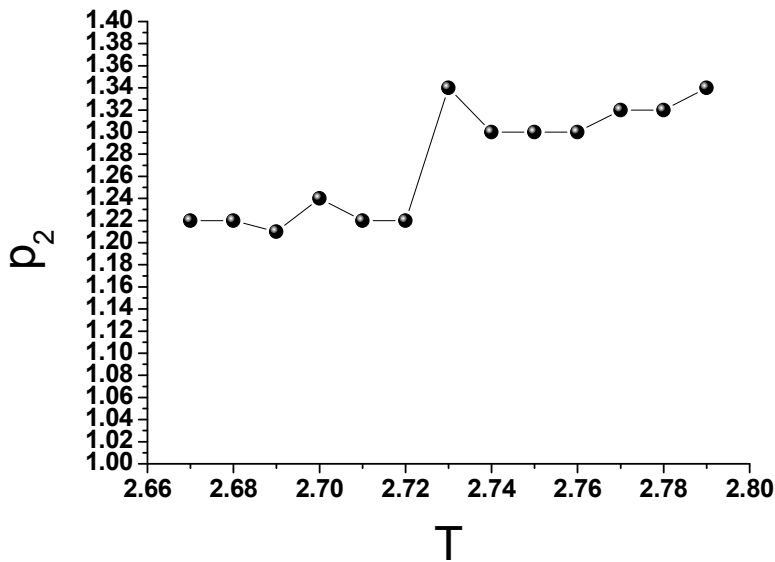

*Fig.3 The $p_2$ values vs the temperature T inside the hysteresis zone lies in the interval between 1, 1.4 . The $p_3$-values are all almost zero . Therefore for all temperatures inside zone the criticality conditions i.e $p_2 \in [1,2)$ and $p_3 \approx 0$, are verified.*

From fig.3 we see that there are two level of values where $p_2$ appear. In the first level after the pseudocritical temperature the mean value is 1.315 which is very close to the 1.33 which is the $p_2$ value for the second order transition in the mean-field Universality class . This is a phenomenon of the degeneration of the critical point which indicate that the critical point has "spread out" in the zone namely it is not localized at only one temperature. At the limit of $L \to \infty$ the critical zone disappears and a single point exists as it is predicted for the critical point of a second order phase transition . The second level has a mean value $p_2 = 1.22$. This result means that the universality class is other. As it is known [**18**] the isothermal critical exponent δ has the value 4.8 for the 3D-Ising universality class and so the correspond $p_2$=1+1/δ =1+1/4.8=1.21. Therefore a hybrid critical state appears . This is a very interesting phenomenon in criticality at finite systems which the reveal cause to chaotic intermittent Dynamics analysis. A qualitative explanation would is that inside zone hysteresis is development a competition between the mean-field mechanism due the action (1) and the geometry of lattice due to the 3D lattice imposing from the 3D-Ising .

## 5. Resonance phenomena inside the zone hysteresis.

As the temperature inside the zone increase we notice from fig. 3 that a jump is accomplished at a different universality class. The existence of this zone of fluctuations is an interesting issue and the study of this zone is a motivation for this work. An interesting question is how this effect is affected with the change of size of Lattice L. So, we repeat the calculations of DW in the fluctuations zone for L=20 and L=30. The results for DW vs T inside this zone is presented in figure 4.

For L=10 a series of spikes appears. For L=30 the spikes have almost vanished. But for an intermediate Lattice with L=20 a resonance phenomenon takes place at temperature 2.72 almost in middle of the zone hysteresis. We could say that this resonance of figure 4 which is connected to the jump of figure 3, since both phenomena take place at the same intermediate temperature 2.72 inside the zone hysteresis , creates a coherence state between the two universality classes. This confirms that indeed in finite systems the critical point degenerates into a series of points within the zone regardless of the universality class to which they belong.

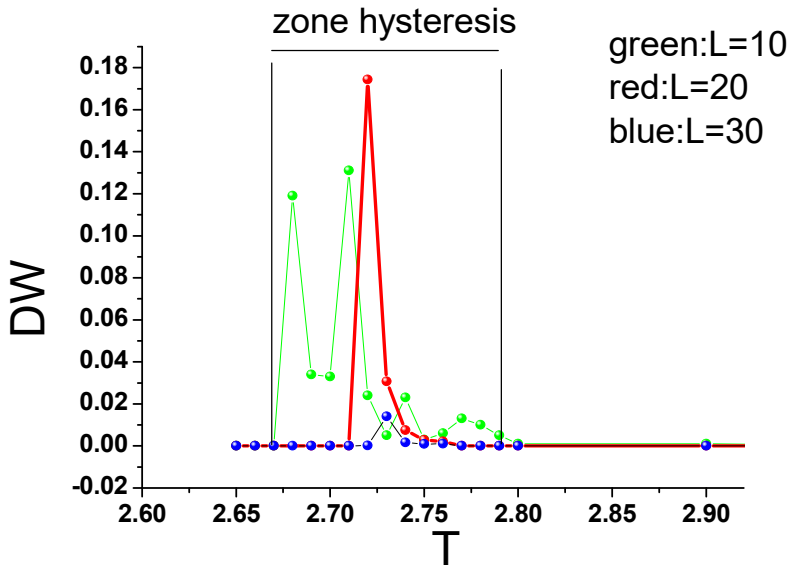

*Fig 4. The DW vs T diagram inside fluctuations zone for lattice with L=10, 20, 30. For L=20 a resonance appears in the middle of zone.*

Such a resonance has been observed in the p-n junction [19] where the critical zone separates the normal rectifier phase in the low frequencies and a full-wave conducting, capacitor like phase in the high frequencies. This resonance accomplished between the mean value of the laminar lengths produced from Dynamics of intermittency vs the frequencies inside the zone. An other example of resonance inside the critical zone is the phase transition of 3D-Ising model where in the middle of the zone a resonance appears between the mean value of the laminar lengths vs the temperatures inside the zone, such as to the p-n junction case [20].

## 6. Other order parameters, the spontaneous symmetry breaking and the tricritical crossover.

In the framework of Landau mean field theory without external field the free energy is given as [**14,21**]:

$$U(\phi) = \frac{1}{2} r_o \phi^2 + \frac{1}{4} u_o \phi^4 \quad (5)$$

Where $\phi$ the order parameter, $u_o > 0$ and $r_o = a_o t = a_o \frac{T-T_c}{T_c}$.

In fig 5 the $U(\phi)$ vs $\phi$ is shown.

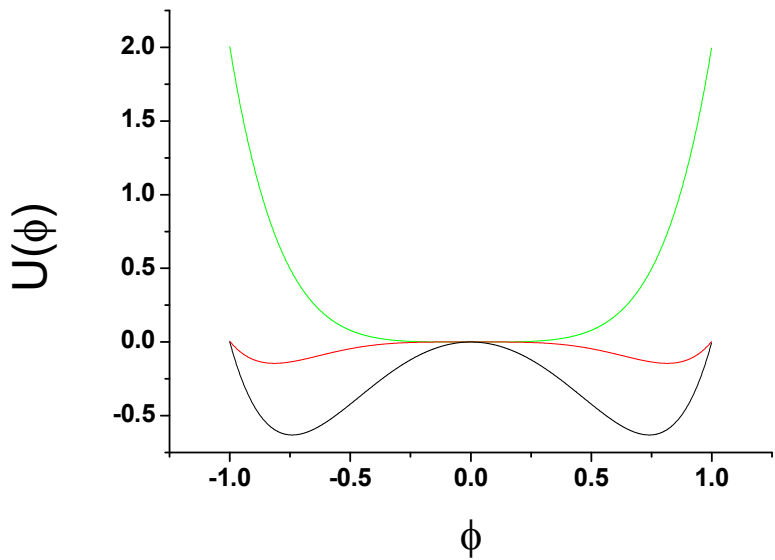

*Fig.5  The Landau free energy in mean field approximation.  The green line refer to the symmetric phase. The red and black lines refer In spontaneous symmetry breaking phenomenon (SSB ) where $r_o < 0, u_o > 0$ .*

According to SSB when the temperature drops under the critical value then the system can exist in only one ground state , even if it possesses many equivalent ones [14]. The  SSB transient follow the fig5 where the two symmetric ground states corresponds to energy minima  whenever in the distributions of the fluctuations of the order parameter values    two lobes appears  (see below fig.6). The SSB is completed when the two lobes do not communicate anymore between them  [10] . After this the system has selected one from two possible state. There are and other quantities which have the characteristics of order parameter , namely are zero in symmetric phase and non  zero in broken symmetry phase. Two such order parameters  are the  measures of the  components of  vector **M = f1+f2+f3**   in the plane x y where projected the whole spins vectors  from  lattice  sites.  Indeed, given that in the  symmetry phase  the vector **M**  is zero  and its components will zero too while  in phase of symmetry breaking the vector **M** is non zero therefore the same is valid for its components.  In fig 6  we present the distribution of the  $M_x=|M_x|$  values at T=2.72.  In fig.5  in  broken symmetry  phase appears  two  minima of energy.  This mean that in distribution of order parameter expected that two lobes , i.e two maxima,   will appears.

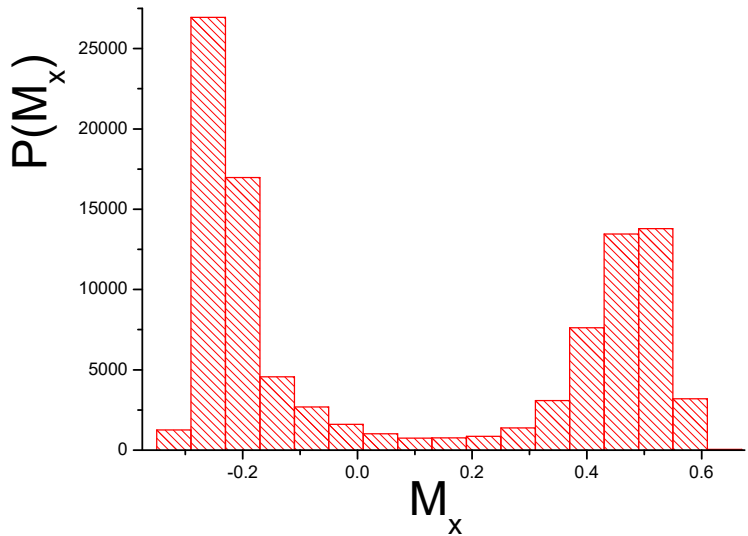

*Fig. 6 The distribution of the measure of the component in axis x . Two lobes are present. For $L \to \infty$ i.e in infinite system , the distribution became symmetric.*

In next figure 7 the timeseries of order parameter $M_x$ for temperature T=2.72 is shown .

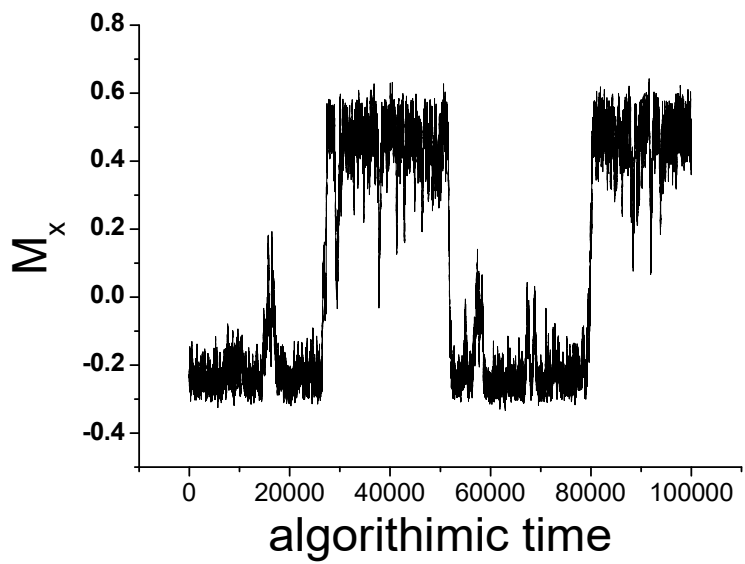

**Fig7. The $M_x$ timeseries shows the dynamics of an on-off intermittency. Such a intermittency we had seen inside the hysteresis zone in phase transition of 3D-Ising model [23] where a second order phase transition takes place.**

We name the up segments of timeseries as up-tms and the down segments as down-tms. The separation point between the two sub-timeseries is found as the smaller $M_x$ value from histogram fig.6. That is $M_x=0.1$. Thus the values which are greater from 0.1 is the up timeseries and the values which are down from 0.1 is the down timeseries. In Fig 7 with red color the line of separation is shown. We will using the MCF, in order to find the Dynamics of each branch.

- We research the laminar region of up timeseries. In fig 7 the end of laminar region estimate at $M_x=0.47$ (line blue) because for this value we find the best power-law for the laminar lengths distribution in the zone of laminar region [0.1,0.47]. In figure 8a this distribution demonstrated.
- We research the laminar region of down timeseries. In fig 7 the end of laminar region estimate at $M_x=-0.23$ (line green) because for this value we find the best power-law for the laminar lengths distribution in the zone of laminar region [-0.23, 0.1]. In figure 8b this distribution demonstrated.

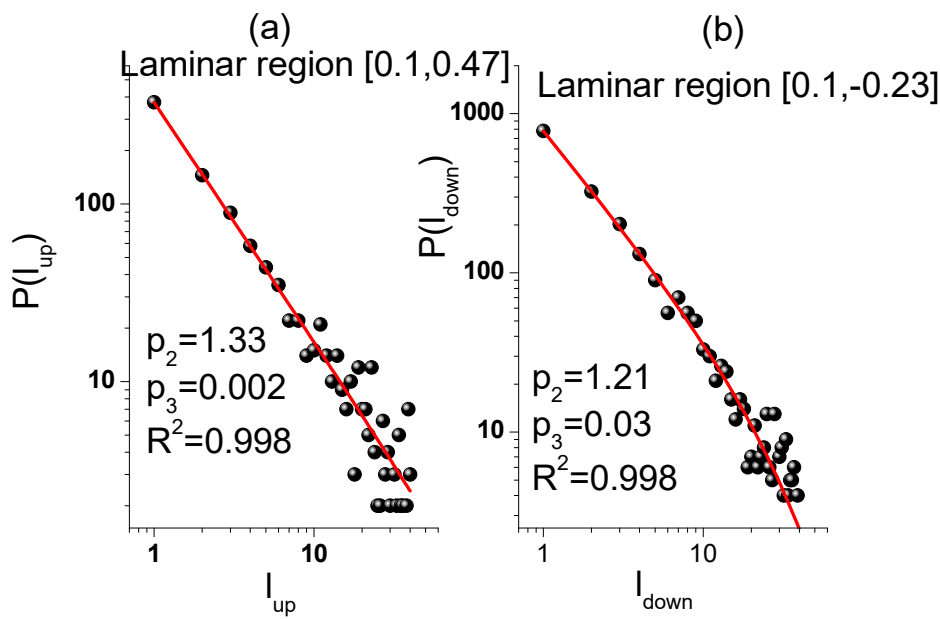

**Fig.8 (a)** The laminar lengths distribution from the up timeseries inside the laminar region [0.1,0.47]. The result is a perfect power-law ($p_3=0.002$) with scaling exponent $p_2=1.33$. **(b)** The laminar lengths distribution from down timeseries inside the laminar region [-0.23,0.1]. The result is very close ($p_3=0.03$) to a power-law with scaling exponent $p_2=1.21$.

The results produced by MCF analysis are very important because these are in absolute agreement with the results produced by MCF in the case of order parameter M. Indeed, the one branch of the order parameter $M_x$ show that the phase transition is second order and belong in the mean field universality class while the other branch of the order parameter $M_x$ show that the phase transition is second order and belong in the 3D Ising universality class. We conclude that the coexistence of the two universality class appears not only with order parameter M but this coexistence appears in the one component too. In other words the MCF when applied in a component of the order parameter M verified the same results when MCF applied in the order parameter M. In order to this happened the Z(3) spin model produce the on-off intermittent chaotic Dynamic in the one of its component.

Let investigate now what happens with the other component i.e the order parameter $M_y$. The Ginzburg-Landau (G-L) free energy with the addition of the $\phi^6$ term is written [**14,21**] as :

$$U(\phi) = \frac{1}{2} r_o \phi^2 + \frac{1}{4} u_o \phi^4 + \frac{1}{6} c_o \phi^6 \quad (6)$$

where $c_o > 0$. An interesting behavior is demonstrated at the region of parameter space where $r_o, c_o > 0$ are positive but $u_o$ changes its sign and becomes negative. In figs. 9b,c,d the tricritical crossover is depicted in diagrams of the G-L free energy (eq.6) vs order parameter $\phi$ .

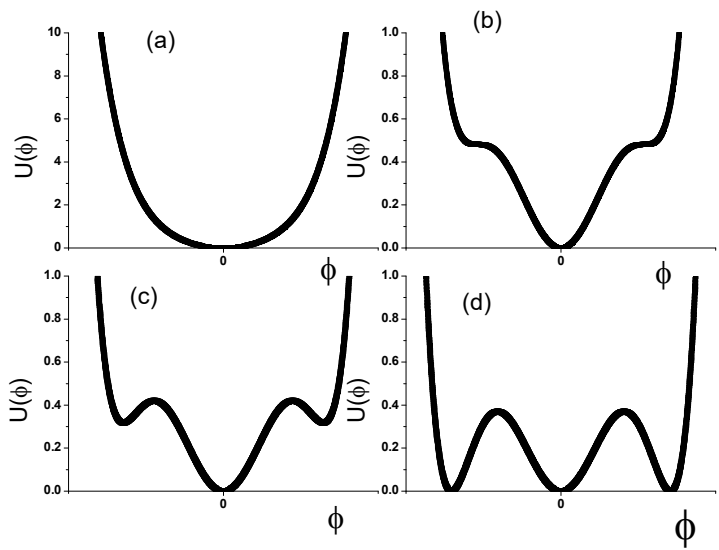

*Figure 9 (a) The free energy of symmetric phase . (b) The beginning of energy deformation. (c) The crossover as metastable phase. (d) The free energy of the first-order phase transition.*

From fig. 9d we see three equal energy minima. This mean that the order parameter in the first order phase transition is in degenerated state because three values in minima positions correspond in the same energy value i.e $U(\phi)=0$. So in the first order phase transitions a discontinuous appears. This is in agreement with the basic property of a first order phase transition which is the abrupt change of order parameter. In figure 9c a metastable phase in the crossover procedure is appears. So the first order phase transition does not have yet completed. Thus we could characterized this state (fig. 9c) as a "weak" first order phase transition. In fig10 we present the distribution of values of the other component of the M, i.e the measure $M_y$ order parameter.

.

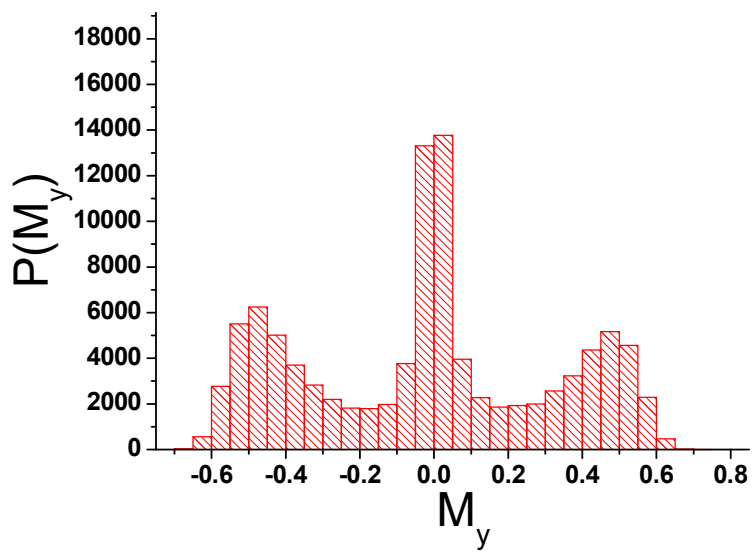

.

*Fig.10 The distribution of the order parameter $M_y$ component during the tricritical crossover appears in the temperature 2.72.*

Comparing the diagram of energy 9c with the distribution in fig10 we see that the central minimum in energy corresponds in the central greater lobe and the two symmetrical local minima of energy correspond to the two smaller lobes. Unfortunately, because the scaling law does not valid during the metastable phase a confirmation through the MCF, as in the case of Mx does not could to become.

Given that where the minima of the energy diagram ( fig9c) corresponds to the maxima of the distribution diagram (fig10) we could take a qualitative figure for energy 9c inversing the distribution diagram (fig.10). Then we take the fig .11 which confirm that the order parameter My corresponds to crossover energy so present a weak first order phase transition.

.

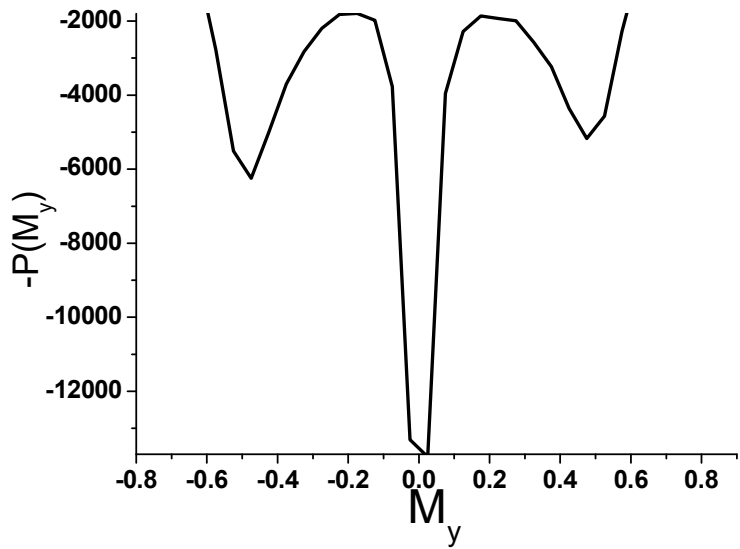

*Fig.11.  We have inverse the fig 10  in order to the lobes of fig.10  became  energy minima of  fig9c. Thus the timeseries data for order parameter My  are in  agreement to a quality description of  a weak first order transition.*

Thus both basic ways  removal from symmetric phase  through the second order phase transition ( SSB)  as well as first order phase transition ( tricritical crossover) appears in a finite Z(3) spin system.

## 7. Conclusions -Discussion

For the Z(3) spin systems where the action is quadratic (eq.1)  we confirm  through the phase diagram of the appropriately defined order parameter and the intermittency  time series analysis   of the order parameter fluctuations through the MCF,  that this is a second order phase transition . The new result is that beyond the continuous phase transition appears and  a weak first order phase transition if the order parameter considered  one (e.g $M_y$)  , from  two  component of the order parameter M.  In the other hand in  the other component (e.g $M_x$) appears  a on-off intermittent chaotic Dynamics which permits a second order phase transition which has the unusual property  to belongs in two universality classes i.e  mean-field  and the model of 3D-Ising. For finite size systems, a zone of fluctuations between the phase of the global symmetry and the phase of the broken symmetry appears. Two major phenomena occur inside this zone. First a degeneration of the critical point of the second-order phase transition occurs by resulting to coherences states and second a resonance phenomenon appears which for infinite systems disappears. Thus a complicated behavior appears in a Z(3) spin model in 3d lattice. Therefore, in the future research, is expected that the symmetry Z(5) will present even more complicated and interesting issues . Due to the fact that Z(3) spin symmetry and SU(3) symmetry of the Quantum Chromodynamic (QCD) have the same centre, it is expected from the universality character of critical phenomena that similar properties appear in SU(3) color space . This is an encouraging clue that the study of the classical phase transitions of such numerical models will help us to further understand better the colorless property of baryons especially , nucleon particles.